\documentstyle{article}
\font\cero=cmss10 scaled 1728 \font\uno=cmssbx10 scaled 1200
\begin{document}
\begin{flushleft}
{\cero The Kodama state for topological quantum field theory
beyond instantons} \\[3em]
\end{flushleft} {\sf R. Cartas-Fuentevilla} and \sf J. F. Tlapanco-Lim\'on\\
{\it Instituto de F\'{\i}sica, Universidad Aut\'onoma de Puebla,
Apartado postal J-48 72570 Puebla Pue., M\'exico }\\

Constructing a symplectic structure that preserves the ordinary
symmetries and the topological invariance for topological
Yang-Mills theory, it is shown that the Kodama (Chern-Simons)
state traditionally associated with a topological phase of
unbroken diffeomorphism invariance for instantons, exists actually
for the complete topological sector of the theory. The case of
gravity is briefly discussed.\vspace{1em}
%\vfill

%\noindent PACS numbers: 04.20.Jb, 04.40.Nr\\
%\noindent Running title:  The Kodama state....\\
%\newpage

\noindent {\uno I. Introduction} \vspace{1em}

As is well known Yang-Mills theory in four dimensions formally
admits the so called Chern-Simons wavefunction as an exact zero
energy eigenfunction of the Schr\"{o}dinger equation \cite{1}.
However, such a state is neither normalizable, nor invariant under
CPT. Additionally negative helicity states not only have negative
energy but also negative norm, and therefore the Chern-Simons
state is not admissible as the ground state of the quantum
Yang-Mills theory \cite{2}. A similar situation is found in the
context of loop quantum gravity with the so called Kodama
state\cite{3}, which is the only solution for the quantum Ashtekar
constraints \cite{4}. Despite these properties, it is important to
understand what these intriguing states describe.

 Recently a deep relationship between topological quantum
field theory and those states has been found \cite{5}. In the
special case of instantons, it has been shown that such states
describe topological phases of unbroken diffeomorphism invariance.
The self-duality conditions on the fields associated with
instantons play a key role producing the deformation of the
original actions into topological actions, picking up thus a
topological phase among various ground states of those theories.
Those states turn out to be the only quantum states for the
topological quantum field theories obtained \cite{5,6}. Along the
same lines, a close relationship between self-duality and the
Kodama state is found also for Abelian gauge theory in \cite{7}.
Therefore, it is natural to associate the Chern-Simons and the
Kodama states with the sector corresponding to instantons.

However, in this letter, we shall show that this is not
necessarily the case, and we shall find that the Chern-Simons
state is associated actually with the whole of the topological
sector of the Yang-Mills theory, without invoking the self-duality
conditions for instantons, provided that we start from the
appropriate topological action.

This work is organized as follows. In the next section we give an
outline of the  ordinary Yang-Mills theory and the topological
version. In Section III, a symplectic structure for topological
Yang-Mills(TYM) theory is constructed. In Section IV the
symmetries of this geometrical structure are considered. For
completeness in Section V the ordinary quantum Yang-Mills theory
and the Chern-Simons state are briefly discussed. Using the
symplectic structure previously constructed , in Section VI the
classical and quantum TYM theory are analyzed. It is shown also
that the corresponding quantum Hamiltonian admits the Chern-Simons
wave-function as an eigenfunction with zero energy. The
constraints are considered in Section VII, and we finish in
Section VIII with
some concluding remarks. \\

\noindent {\uno II. Ordinary and  TYM theory} \vspace{1em}

The usual Yang-Mills action is given by
\begin{equation}
     S_{YM} (A) = \alpha \int_{M} {\rm Tr} \ (F \wedge
     ^{\star} F),
\end{equation}
where $\alpha$ is a parameter, $A$ is the gauge connection, and $F
= dA + A \wedge A$ its curvature; $d$ and $\wedge$ correspond to
the exterior derivative and the wedge product on $M$, which we
assume as the four-dimensional Minkowski spacetime. $^{\star}F$ is
the dual of $F$ in the usual sense. Due to this duality operation,
the Yang-Mills action depends on the metric structure on $M$.

The action (1) implies the Yang-Mills equations
\begin{equation}
     d ^{\star} F + [A, ^{\star} F] = 0,
\end{equation}
with the curvature $F$ satisfying the Bianchi identity
\begin{equation}
     dF + [A,F] = 0.
\end{equation}
As well known, when the connection is self-dual or anti-self-dual
$F = \pm ^{\star}F$, the Bianchi identity (3) automatically
implies the Yang-Mills equations (2), and additionally the action
(1) turns out to be a topological action, independent on the
metric of $M$.

Alternatively, one can construct a TYM theory without invoking the
self-duality condition,
\begin{equation}
     S_{TYM} (A) = \beta \int_{M} {\rm Tr} \ (F \wedge F),
\end{equation}
which does not depend on the metric on $M$, however, it does on
the smooth structure of $M$. Thus,
\begin{equation}
     \frac{\delta S_{TYM}}{\delta g_{\mu\nu}} = 0,
\end{equation}
where $g_{\mu\nu}$ are the components of the metric on $M$. The
action (4) implies trivially the equations (3), {\it i.e.} every
gauge connection $A$ is a critical point for this action. The
action (4)
is the subject of the present study. \\

\noindent {\uno III. The symplectic structure for TYM theory}
\vspace{1em}

Following Appendix A in \cite{8}, one can construct from the
action (4) a symplectic structure that preserves all relevant
symmetries of the theory. The variation of the action (4) is given
by
\begin{equation}
     \delta S_{TYM} (A) = \int_{M} {\rm Tr} \ \partial_{\mu} [4\beta
     \widetilde{F}^{\mu\nu} \delta A_{\nu}] d^{4}x - 4 \beta
     \int_{M} {\rm Tr} \ (D_{\mu} \widetilde{F}^{\mu\nu}) \delta
     A_{\nu}d^{4}x,
\end{equation}
where we have displayed explicitly the components in order to make
direct contact with \cite{2,5}. From (6) we identify the Bianchi
identity (3) in terms of components,
\begin{equation}
     D_{\mu} \widetilde{F}^{\mu\nu} \equiv \partial_{\mu}
     \widetilde{F}^{\mu\nu} + [A_{\mu}, \widetilde{F}^{\mu\nu}] =
     0,
\end{equation}
and the argument of the total divergence as a {\it symplectic
potential} for the theory
\begin{equation}
     \Theta^{\mu} \equiv 4 \beta {\rm Tr} \ \widetilde{F}^{\mu\nu}
     \delta A_{\nu},
\end{equation}
whose variations give the integral kernel of the symplectic
structure:
\begin{equation}
     \omega = \int_{\Sigma} 4 \beta {\rm Tr} \ \delta
     (\widetilde{F}^{\mu\nu} \delta A_{\nu}) d \Sigma_{\mu} =
     \int_{\Sigma} 4 \beta {\rm Tr} \ \delta \widetilde{F}^{\mu\nu}
     \delta A_{\nu} d \Sigma_{\mu},
\end{equation}
where $\Sigma$ is a Cauchy hypersurface, and we have considered
that $\delta$ corresponds to the exterior derivative on the phase
space $Z$ of the theory \cite{9}, which is understood as a
submanifold of the {\it kinematic space} (the space of smooth
connections $A$ and curvatures ${F}$), on which the {\it
constrictions} (7) hold. It is important to mention that $Z$ has
not {\it a priori} a symplectic structure, for example (9), but an
action principle is a necessary ingredient for that \cite{10}.

Furthermore, using the first-order variation of Eq.\ (7) given by
\[
     D_{\mu} \ \delta \widetilde{F}^{\alpha\mu} + [\delta A_{\mu},
     \widetilde{F}^{\alpha\mu}] = 0,
\]
we find that the integral kernel of $\omega$ is covariantly
conserved, $\partial_{\mu} \ ({\rm Tr} \ \delta
\widetilde{F}^{\mu\nu} \delta A_{\nu}) = 0$, which makes $\omega$
independent on the choice of $\Sigma$. \\

\noindent {\uno IV. Symmetries of $\omega$} \vspace{1em}

Let us show that $\omega$ retains the symmetries of the
topological action (4). Since $\delta \widetilde{F}^{\mu\nu}$ and
$\delta A_{\nu}$ transform homogeneously under the infinitesimal
gauge transformation
\begin{equation}
     A_{\mu} \rightarrow A_{\mu} + D_{\mu} \varepsilon,
\end{equation}
$\omega$ is gauge invariant. Furthermore, let $\widehat{Z}$ be the
phase space $Z$ modulo the action of the symmetry group, and let
us show that $\omega$ has not components tangent to the gauge
directions, which are specified by
\begin{equation}
     \delta A_{\mu}^{'} \rightarrow \delta A_{\mu} + D_{\mu}
     \varepsilon;
\end{equation}
$\omega$ will undergo the transformation
\begin{equation}
     \omega^{'} = \omega + \int_{\Sigma} \partial_{\mu} {\rm Tr}
     [\widetilde{F}^{\mu\nu} \varepsilon ]\ d \Sigma_{\nu},
\end{equation}
where Eqs.\ (7) have been considered; hence, Eq.\ (12) shows, for
fields with compact support, that $\omega$ is a gauge invariant
symplectic structure on $\widehat{Z}$.

On the other hand, the action (4) possesses the topological
invariance in the sense of (5). It is straightforward to show that
$\omega$ is a topological invariant in the same sense, since its
expression (9) in terms of components shows that $\omega$ does not
depend on the metric of $M$:
\begin{equation}
     \omega = \int_{\Sigma} 4\beta \ {\rm Tr} \
     \epsilon^{\mu\nu\alpha\beta} \ \delta F_{\alpha\beta} \ \delta
     A_{\nu} \ d \Sigma_{\mu},
\end{equation}
which can be expressed in a compact form as
\begin{equation}
     \omega = \int_{\Sigma} 4\beta \ {\rm Tr} \ \delta F \wedge \
     \delta A,
\end{equation}
which shows clearly the independence of $\omega$ on the metric
structure of $M$, such as the action itself in (4). Thus
\begin{equation}
     \frac{\delta\omega}{\delta g_{\mu\nu}} = 0.
\end{equation}
However, as well known, the topological action (4) is not
invariant under the parity operation $P$, and thus $CP$ and $CPT$
are not symmetries for such an action. From the expressions (9) or
(14), it is easy to note that $\omega$ inherits this property.
\\

\noindent {\uno V. The Chern-Simons state for ordinary YM theory}
\vspace{1em}

For completeness we give an outline of what is known about the
Chern-Simons state in conventional YM theory.

 The quantum Hamiltonian for Yang-Mills theory can be
obtained from the classical expression for the energy,
\begin{equation}
     H = \frac{1}{2g^{2}} \int d^{3} x \ {\rm Tr} \ (E^{2} + B^{2})
     = \frac{1}{2} \int d^{3} x \ {\rm Tr} \ \big( -g^{2}
     \frac{\delta^{2}}{\delta A(x)^{2}} + \frac{1}{g^{2}} B^{2} \big),
\end{equation}
where $g$ is the gauge coupling, $E_{i} = F_{0i}$, $B_{i} =
\frac{1}{2} \epsilon_{ijk} F_{jk}$, and quantum mechanically the
canonical momentum $\frac{E}{g^{2}}$ becomes
$-i\frac{\delta}{\delta A}$:
\begin{equation}
\frac{F_{0i}}{g^{2}} \rightarrow -i \frac{\delta}{\delta A_{i}}.
\end{equation}
The Chern-Simons functional
\begin{equation}
     I = \frac{1}{4\pi} \int d^{3} x \ {\rm Tr} \ ( \epsilon^{ijk} \big( A_{i}
     \partial_{j} A_{k} + \frac{2}{3} A_{i}A_{j}A_{k} \big) ),
\end{equation}
for which $\frac{\delta I}{\delta A} = \frac{B}{2\pi}$, allows to
construct the wave-function
\begin{equation}
     \psi (A) = exp \big[ \big( \frac{2\pi}{g^{2}} \big) I (A) \big],
\end{equation}
satisfying
\begin{equation}
     (E + iB) \psi = i \big( -g^{2} \frac{\delta}{\delta A} +
     B \big) \psi= 0,
\end{equation}
and thus $\psi$ corresponds to an eigenstate of the Hamiltonian
(16) with zero energy. \\

\noindent {\uno VI. Classical and Quantum Hamiltonian for TYM
theory and the Chern-Simons state} \vspace{1em}

We can obtain the physical content of the symplectic structure
$\omega$ following \cite{9}, and to consider the contraction of
$\omega$ with the phase space vector field $V$ corresponding to
the translation by a constant spacetime vector
$\varepsilon^{\mu}$. Considering that
\[
     V \rfloor \delta A^{\mu} = \varepsilon_{\beta} \
     F^{\beta\mu},
\]
we have
\[
     V \rfloor \delta \widetilde{F}^{\alpha\mu} =
     \varepsilon^{\beta} D_{\beta} \widetilde{F}^{\alpha\mu},
\]
and hence, using Eq.\ (9),
\[
     V \rfloor \omega = \varepsilon_{\beta} \int d \Sigma_{\alpha}
     {\rm Tr} \ \delta (-4\beta) \big[ F^{(\beta}{_{\mu}}
     \widetilde{F}^{\alpha )\mu} - \frac{1}{4} n^{\alpha\beta} \
     F_{\lambda\mu} \widetilde{F}^{\lambda\mu} \big],
\]
modulo total divergences. Therefore, from the above equation we
can identify the (symmetric and gauge-invariant) energy-momentum
tensor,
\begin{equation}
     T^{\mu\nu} = - {\rm Tr} \ 4\beta \big( F ^{(\mu} {_{\alpha}}
     \widetilde{F}^{\nu )\alpha} - \frac{1}{4} g^{\mu\nu}
     \widetilde{F}^{\mu\nu} F_{\mu\nu} \big),
\end{equation}
which is classically zero, as expected for a topological action,
in concordance with Eq.\ (5). From Eq.\ (21) we can identify the
energy density for the theory,
\[
     T^{00} = -2{\rm Tr} \ \beta \big( F_{0i} \widetilde{F}^{0i} - \frac{1}{2}
     F_{ij} \widetilde{F}^{ij} \big) = {\rm Tr} \ F_{0i} (2\beta
     \widetilde{F}_{0i} - \beta \ \epsilon_{ijk} F_{jk}) = {\rm Tr} \ F_{0i}
     (\pi_{i} - \beta \ \epsilon_{ijk} F_{jk}),
\]
where $\widetilde{F}^{ij} = \epsilon^{ijk} F_{0k}$, and $\pi_{i}
\equiv 2\beta \widetilde{F}_{0i}$; hence the classical Hamiltonian
is given by
\begin{equation}
     H = \int_{\Sigma} d \Sigma  {\rm Tr} \ F_{0i} (\pi_{i} - \beta \epsilon_{ijk}
     F_{jk}).
\end{equation}
The idea is to use the symplectic structure $\omega$ constructed
previously in Section III for obtaining from (22) the
corresponding quantum Hamiltonian. Therefore, considering that
$d\Sigma_{\mu}$ is a time-like vector field, we can obtain in
particular the following (non-covariant) description of the phase
space,
\begin{equation}
     \omega = \int_{\Sigma}  \ 4\beta \ {\rm Tr} \ (\delta \widetilde{F}^{0i}
     \wedge \delta A_{i}),
\end{equation}
in order to make contact with \cite{2,5}. This expression for the
symplectic structure will allow us  to work in the {\it temporal
gauge} $A_{0} = 0$. Furthermore, Eq.\ (23) shows explicitly that
the canonical variables for TYM theory are given by $(2\beta)
\widetilde{F}^{0i}$ and $A_{i}$. Therefore, we have the
classical-quantum correspondence,
\begin{equation}
     (2\beta) \widetilde{F}_{0i} \rightarrow i \frac{\delta}{\delta
     A_{i}}.
\end{equation}
Making a comparison with Eq.\ (17), we see that the canonical
momentum for TYM theory is the dual of that for conventional YM
theory. In the case of self-dual fields, the correspondence (24)
reduces to that in Eq.\ (17), and similarly the expressions (23),
(22), and (21) reduce to the corresponding expressions given in
Section V, in the same sense that the topological action (4)
becomes that in Eq.\ (1). Thus we have, as a particular scenario,
the self-dual case. However, we avoid at all, as already mentioned
in the introduction, the {\it temptation} of imposing such a
condition, and we shall prove that the Chern-Simons state is an
eigenstate with zero energy for TYM theory without invoking the
self-dual condition.

In this manner, considering Eq.\ (24), the quantum Hamiltonian
will be given by
\begin{equation}
     H_{Q} = \int_{\Sigma} d \Sigma \ {\rm Tr} \  F_{0i} \big( i\frac{\delta}{\delta
     A_{i}} - \beta \ \epsilon_{ijk} F_{jk} \big).
\end{equation}
In the temporal gauge $A_{0} = 0$, it is easy to show that
$[F_{0i}, \frac{\delta}{\delta A_{i}}] = 0$, and thus we have no
ordering ambiguity in the quantum Hamiltonian (25). Thus, any wave
function $\psi$ satisfying
\begin{equation}
     \big( i \frac{\delta}{\delta A_{i}} - \beta \ \epsilon_{ijk}
     F_{jk} \big) \psi = 0,
\end{equation}
will correspond to a state of zero energy for the Hamiltonian
(25). Therefore, the solution for Eq.\ (26) is given by
\begin{equation}
     \psi (A) = e^{-4\pi i\beta I(A)},
\end{equation}
with $I(A)$ given in Eq.\ (18). Equation (27) is essentially the
Chern-Simons wave function (19), and we conclude then that the
Chern-Simons state corresponds strictly to a (topological) state
of the TYM theory, without invoking the self-duality condition on
the fields. Thus, such a state is associated with the topological
phase of unbroken diffeomorphism invariance of the
complete topological sector.\\

\noindent {\uno VII. The constraints} \vspace{1em}

Equations (7) imply that
\begin{equation}
     D_{i} \widetilde{F}^{i0} = 0,
\end{equation}
which must be considered as a constraint on the quantum states, in
particular on the Chern-Simons state. Considering the
correspondence (24) for TYM theory, such a constraint reads
\begin{equation}
     D_{i} \frac{\delta}{\delta A_{i}} \psi = 0,
\end{equation}
which ensures that $\psi$ is unchanged under {\it small} gauge
transformations \cite{1}. Note that for self-dual fields, Eq.\
(28) reduces to the Gauss law for conventional YM theory, and (29)
will represent the usual Gauss law constraint. However, as we have
shown, the constraint (29) and its relationship with invariance of
$\psi$ under small gauge transformations exist in the whole of the
topological sector of the theory and, in particular, for
instantons. In a more familiar way, Eq.\ (28) corresponds
classically to the equation $D_{i} B^{i} = 0$, and thus we can
identify it as the generator of gauge symmetries for TYM theory at
a quantum level in accordance with Eq.\ (29).

On the other hand, within the Dirac quantization scheme, the
correspondence (24) gives rise to the {\it primary} constraint
\begin{equation}
     i\frac{\delta}{\delta A_{i}} - 2 \beta \widetilde{F}_{0i} = i
     \frac{\delta}{\delta A_{i}} - \beta \ \epsilon_{ijk} \ F_{jk}
     \approx 0,
\end{equation}
which is of first-class, and thus any physical quantum state
$\psi$ must satisfy
\begin{equation}
     \big( i\frac{\delta}{\delta A_{i}} - \beta \ \epsilon_{ijk} \
     F_{jk} \big) \psi = 0,
\end{equation}
that corresponds exactly to Eq.\ (26) for the Chern-Simons state.
In this sense, the Hamiltonian (25) is purely a combination of the
constraints (31), with ${F}_{0i}$, the dual of the canonical
momentum, playing the role of a Lagrange multiplier field. \\

\noindent {\uno VIII. Concluding remarks} \vspace{1em}

It is important to remark at this point the results obtained. The
starting point is the topological Yang-Mills theory, which has as
{\it equations of motion} the Bianchi identities, and as {\it
solution space} the complete space of gauge connections. On this
solution space the topological action defines a symplectic
structure, and the corresponding Hamiltonian admits the
Chern-Simons wavefunction as a zero energy eigenfuction. Hence,
the Chern-Simons state exists for the complete topological sector
of the theory, and in order to establish its existence, neither
the self-dual condition nor the Yang-Mills equations are required.
The Bianchi identity becomes the generator of gauge symmetries at
quantum level. As a particular case, the topological phases of
Yang-Mills theory can be obtained invoking self-duality.

The subject of references \cite{2,5} is focused on the question
{\it why the Chern-Simons state exists}. The present results
suggest that this question should be translated beyond the
instantons sector, {\it i.e.} to the complete topological sector
of the theory. Furthermore, the problem of the normalizability of
this state under an appropriate inner product must be also
reformulated in the scheme of the TYM theory, since the inner
product used in the ordinary field theory seems to fail in the
topological scheme.

 Although instantons play a vital role in
fundamental aspects of field theories, it is of vital importance
also to clarify those aspects that are not necessarily related
with such topological objects. In this sense, the problem of
studying the sectors beyond the instantons even persists; it is
possible that those aspects normally associated with instantons,
are actually related with the whole of the topological sectors.

Following the arguments given in the present treatment for TYM
theory, it is possible that in the context of loop quantum gravity
similar results can be obtained; however, this will be a problem
for future works.

\begin{center}
{\uno ACKNOWLEDGMENTS}
\end{center}
\vspace{1em}

This work was supported by the Sistema
Nacional de Investigadores and Conacyt (M\'{e}xico).\\

\end{document}